\documentclass[12pt]{article}
\usepackage{amsmath}

\input{tcilatex}

\begin{document}

\begin{titlepage}
\begin{flushright}
PUPT-2244\\
hep-th/yymmnnn
\end{flushright}
\vspace{7 mm}
\begin{center}
\huge{De Sitter Space and Eternity}
\end{center}
\vspace{10 mm}
\begin{center}
{\large
A.M.~Polyakov\\
}
\vspace{3mm}
Joseph Henry Laboratories\\
Princeton University\\
Princeton, New Jersey 08544
\end{center}
\vspace{7mm}
\begin{center}
{\large Abstract}
\end{center}
\noindent
This paper explores infrared quantum effects in the de Sitter space.
The notion of "eternal manifolds" is introduced and it is shown that in most
cases the de Sitter space doesn't belong to this class. It is unstable under
small perturbations which may cause a breakdown of the de Sitter symmetry.
The de Sitter string sigma model is discussed. It is argued that
the gauge theory at the complex coupling is dual to the matrix elements of vertex
operators in the de Sitter
space, taken between the Bunch - Davies vacuum and the "out" state without particles.
The described infrared effects are likely to screen away the cosmological constant.
\vspace{7mm}
\begin{flushleft}
September 2007
\end{flushleft}
\end{titlepage}\bigskip\ 

\section{\protect\bigskip Introduction.}

Many years ago I conjectured \cite{p81} that the cosmological constant may
be screened by the infrared fluctuations of the metric, much like the
electric charge in quantum electrodynamics. This effect, if it exists, must
be non-perturbative, related to the fluctuations of the metric $g_{\mu\nu}$
near zero, and not near a classical background value. In the time- dependent
picture the screening is equivalent to the instability of space- times with
constant curvature. In this picture the initially present curvature is
gradually decaying. It is important, therefore to find out, whether the de
Sitter space carries the infrared seeds of its own destruction. This is the
topic of the present article.

It is not straightforward to find an appropriate framework for discussing
this question. The notion of fields and particles in the curved Lorentzian
backgrounds is ambiguous and the Hamiltonian may not exist at all. We have
to formulate a principle which describes quantum field theories in these
circumstances. With this goal in mind we will postulate that a free particle
on a stable manifold must propagate with an amplitude%
\begin{equation}
G_{++}(x,x^{\prime })=\sum_{(P_{xx^{\prime }})}e^{-imL(P_{xx^{\prime }})}
\end{equation}%
where the sum goes over all paths belonging to the manifold and connecting
the points $x$ and $x^{\prime }$ , $L(P)$ is the length of the path , $m$ is
the mass of the particle and the subscripts will be explained later. When we
consider interacting particles, the Feynman prescription is to draw the
diagrams and to integrate the position of the vertices over the whole
space-time.

That seems to contradict causality, since the observables at a given moment
of time seem to depend on the interaction in the future. The resolution of
this puzzle is well known. The use of Feynman's diagrams presupposes that
the vacuum is stable , that nothing new will ever happen. This assumption
restores the causality and leads to $CPT$ symmetry. In this case the
manifold is eternal and the future is being prefixed. We define (
perturbatively ) "\textit{eternal manifolds} " as the ones for which the
Feynman rules are valid and the vacuum loops do not contain an imaginary
parts.

If, however, the vacuum is unstable, the picture is quite different. To
calculate the present we must eliminate all the signals from the future. The
way to do it was discovered by Schwinger \cite{schw} and developed by a
number of people \cite{keld} . According to \cite{schw} , we must double the
manifold and consider the two copies $(+/-)$ with the chronological ordering
from $-\infty$ to $\infty$ on the first sheet and back on the second. The
vertices of Feynman's diagrams are now labeled by their position $x$ and an
Ising variable $\sigma=+/-$ indicating the choice of the sheet.

The \textit{eternity test }is the condition that all diagrams in which a
particle changes the sheet ( thus containing the $G_{+-}$ Green function)
are zero. A slightly different way to put it is to require that all diagrams
containing "spiders" - a $(-)$ point surrounded by $(+)$ -s only are zero. A
spider represents spontaneous particle creation from the vacuum. If this is
forbidden we return to the ordinary Feynman's diagrams. We will explore this
condition below and find that in the de Sitter space in the certain cases it
is violated. The next step will be to understand this from the point of view
of gauge/strings duality.

If de Sitter space turns out to be intrinsically unstable, it must be
described \textit{\ by a non-unitary conformal field theory}. A natural
candidate for such a theory would be a Yang -Mills theory with complex
coupling obtained by the analytic continuation from negative to positive
curvature. We also should expect to find a description in terms of the
world-sheet sigma model with the de Sitter space as a target.

These statements are puzzling since the de Sitter space has been
investigated in many works and no instability of the so called Bunch -
Davies vacuum was ever found (see \cite{birr}, \cite{stro} for the excellent
reviews). My assertion here is that neither this nor any other vacuum passes
the "eternity test". It is perhaps appropriate to explain the meaning of
this test, leaving technical details (discussed below).

Let us consider the Bunch - Davies vacuum for a scalar field $\varphi.$This
field in the de Sitter space with the standard metric $ds^{2}=\cosh
^{2}td\Omega^{2}-dt^{2}$ can be expanded in partial waves. Each partial wave 
$\varphi_{m}$ ( $m$ being angular momentum, see below ) can be further
expanded as $\varphi_{m}=a_{m}\phi_{m}(t)+a_{m}^{\dag}\phi_{m}^{\ast}$ ,
where $\phi-$ s are the appropriate solutions of the Klein-Gordon equation (
27) and $a,a^{\dag}$ are the creation and annihilation operators. The Bunch-
Davies (aka Hartle- Hawking, aka Euclidean) vacuum $E$ is defined as $%
a|E\rangle=0.$ There are also two other vacua $|\pm\rangle$, defined with
the modes without positive frequencies in the future or negative in the past
(we will also call them "in" and "out" states when it is not ambiguous).
These are the states without particles at the past / future infinity
respectively. They have the feature $|\langle E|\pm\rangle|<1,$meaning that
in the Euclidean vacuum particles are created. By itself this is not yet an
instability. One might think that the Euclidean vacuum is eternal , but
viewed from infinity, contains particles. Until we include interactions this
is a possible point of view.

Let us now take, say, the $\varphi^{4}$ interaction and turn it on
adiabatically and then turn it off in the same way. In the usual field
theory the vacuum will acquire a phase after this procedure but otherwise
will remain unchanged. Not so in the de Sitter space. From the formulae
given below it follows that $\langle E|T\exp$-i$\int\lambda(\epsilon
t)\varphi^{4}(dx)|E\rangle\rightarrow0$ as $\epsilon\rightarrow0.$In other
words, with any interaction $|E\rangle_{out}$ is orthogonal to $%
|E\rangle_{in}$ . Technically this "\textit{adiabatic catastrophe}" occurs
since we can evaluate the above matrix element by the usual Feynman
diagrams, and, as we will see, they contain very strong infrared divergences.

One can try to study instabilities in a different way, by looking at the $%
in/in$ matrix elements and using the Schwinger prescriptions. This is a kind
of a \textit{"laissez - faire"} approach when we let the system develop by
itself. In this framework the instability means that small perturbations
breaking the de Sitter symmetry grow with time, while the divergences ,
requiring the IR cut-off (like $\epsilon $ in the above formula) go away. It
is reasonable to expect that as a result of these infrared phenomena, a
small perturbations of the metric and other fields will be amplified with
time, destroying the de Sitter symmetry. This can be viewed as a kind of
spontaneous symmetry breaking ( with large value of time playing the role of
the large volume in statistical mechanics). We must stress, however, that
there is still no complete proof of the instability in the Schwinger
framework. It remains to be proved that in this case the vacuum polarization
will lead to some kind of gravitational instability. Below we will give some
arguments in favour of this statement. A closely related phenomenon happens
in the Starobinski model \cite{starob} where the instability occurs due to
the trace anomaly.

Another limitation of the present work is that it treats only interacting
scalar fields and leaves aside the tensor structure and gauge fixing of the
gravitational oscillations. Hopefully the qualitative effects we are after
will still be present in the full theory.The "dS/CFT correspondence" has
been discussed in the papers \cite{stro1},\cite{witt},\cite{malda}. While
the approaches of \cite{stro1},\cite{witt} are quite different from the
present paper, the ideas of \cite{malda} go in the same direction . Still,
our conclusions below seem new. Many formulae for the various propagators
used below must be well known to the experts in the field. Infrared
divergences have also been investigated before \cite{wood} \cite{weinb}, but
we add some new results and approaches. The relation between infrared
divergences and stability and the relation to the non-unitary gauge theory
was discussed in \cite{p06} on a qualitative level; see also a recent
discussion of the Euclidean infrared problems \cite{jack} . The particle
production , instabilities and back reaction have been discussed from the
various points of view in the papers \cite{motta} ,\cite{ant} , \cite{klu}
and references therein. A runaway particle production was studied in an
interesting paper \cite{myr}

\section{The composition principle and the Unruh detector}

The formula (1) implies the asymptotic condition which uniquely fixes the
propagator%
\begin{equation}
G(x,x^{\prime})\sim e^{-imL(x,x^{\prime})}
\end{equation}
where $L(x,x^{\prime})\rightarrow\infty$ being the geodesic time-like
length. Another way of putting it is to select the Green function of the
Klein - Gordon equation which, for the time-like separations is analytic in
mass if $\func{Im}m<0.$

It is important to realize that the generic propagator would contain a
superposition $Ae^{imL}+Be^{-imL}.$ This propagator would violate the
following composition principle which lies at the foundation of quantum
field theory. We should be able to glue together two paths and obtain a
single one. This condition is equivalent to unitarity in the ordinary
quantum field theory. In the classical limit this gluing procedure is
expressed by the relation $L(x,x^{\prime })=L(x,y)+L(y,x^{\prime })$ where $%
y $ must extremize this expression. In the quantum case one has to account
for the entropy on the world line (or the world sheet in the case of
strings) and the relation is 
\begin{equation}
\int dyG(x,y)G(y,x^{\prime })=\sum_{(P_{xx^{\prime }})}L(P)e^{-imL(P)}\sim 
\frac{\partial }{\partial m}G(x,x^{\prime })
\end{equation}%
As a side remark, let us notice that in the case of strings the extra factor 
$L$ in this formula is replaced by the free energy of the self-avoiding path
belonging to the surface. These gluing relations in the flat space play the
role of unitarity conditions for the corresponding field theory. It seems
reasonable to postulate that the "right" propagators must respect these
conditions in the general case. If the two exponents are present, we would
get in the asymptotic the term $\widetilde{L(}x,x^{\prime
})=L(x,y)-L(y,x^{\prime }),$ which , being minimized with respect to y,
doesn't reproduce the left hand side.

The above condition has a curious relation to the behaviour of the Unruh
detector . As well known \cite{birr} , the probability for absorbing energy $%
\varepsilon $ by this detector is given by%
\begin{equation*}
w(\varepsilon )\sim \int ds_{1}ds_{2}e^{-i\varepsilon
(s_{1}-s_{2})}G(x(s_{1}),x(s_{2}))
\end{equation*}%
where $x(s)$ is a trajectory of the detector and $s$ is its proper time. Let
us assume that the trajectory is a geodesic. In this case $G\sim
Ae^{-im(s_{1}-s_{2})}+Be^{im(s_{1}-s_{2})}.$ We see that the $B$ coefficient
,which violates the composition, simultaneously defines the absorbtion
probability of the inertial detector. If it is non-zero, \textit{the
inertial detector can borrow energy from the vacuum indefinitely !. }It is
not surprising that such a vacuum won't live long.

Let us now turn to the de Sitter space. The standard approach uses the
propagators obtained by analytic continuation from the sphere (which define
the Bunch- Davies or Euclidean vacuum). Let us have a look at them, using
for simplicity the 2d sphere. The Green function is given by 
\begin{equation}
G(n,n^{\prime })=\sum_{l=0}^{\infty }\frac{(2l+1)P_{l}(n\cdot n^{\prime })}{%
l(l+1)+M^{2}}=\frac{\pi }{\sin \pi \nu }P_{\nu }(-n\cdot n^{\prime })
\end{equation}%
here $n$ is a unit vector, $\nu $ is defined by a relation $M^{2}=-\nu (\nu
+1)$ and the above formula is obvious since the left hand side has the same
poles and residues as the right one. It also has the expected logarithmic
singularity at the coinciding points $(n\cdot n^{\prime })=1.$ On a sphere
it satisfies the composition principle 
\begin{equation}
\frac{1}{2\nu +1}\frac{\partial }{\partial \nu }(\frac{1}{\sin \pi \nu }%
P_{\nu }(-n\cdot n^{\prime }))=\frac{\pi }{\sin ^{2}\pi \nu }\int
dn_{1}P_{\nu }(-n\cdot n_{1})P(-n_{1}\cdot n^{\prime })
\end{equation}%
For the future references, let us point out that in the $d+1$ dimensional
case the only changes in these formulae are 
\begin{equation}
P_{\nu }(z)\Rightarrow C_{\nu }^{\frac{d}{2}}(z);l(l+1)\Rightarrow
l(l+d);2l+1\Rightarrow 2l+d
\end{equation}%
(where $C$ are the Gegenbauer polynomials). We will obtain the $dS/AdS$
spaces by analytic continuation from the sphere. But the procedure is
nontrivial and is discussed in the next section

\section{ Analytic connections of various spaces.}

The de Sitter space (or imaginary Lobachevsky space) is obtained from a
sphere by an analytic continuation $n_{0}\Rightarrow in_{0}$ and forms a one
sheeted hyperboloid%
\begin{equation}
n_{k}^{2}-n_{0}^{2}=1
\end{equation}
The metric has the form $ds^{2}=(dn_{k})^{2}-(dn_{0})^{2}.$Since even
complex change of variables doesn't change the scalar curvature, it remains
constant and positive. The distance $L$ between two points is given by the
formula $\cosh L=(n\cdot n^{\prime})$ and could be both time-like (real $L$)
and space-like (imaginary $L$). The Klein-Gordon (or Laplace) equation,
defining a propagator is also unchanged as we make the analytic
continuation. Therefore we can obtain propagators on the de Sitter space
from the propagator on a sphere by the simple substitution. The only problem
is that a physical propagator in one case may be unphysical in another.

The Euclidean vacuum is defined by the propagator ( 4) in which we
substitute $(n\cdot n^{\prime })=(n_{k}n_{k}^{\prime })-n_{0}n_{0}^{\prime }$%
. This is a nice, well defined function with the expected singularity.
However, it doesn't satisfy our additivity requirement. Indeed, in the limit
of large $L$ we get the following asymptotics , using the formulae $P_{\nu
}(z)\sim Az^{\nu }+Bz^{-\nu -1}$ (at large $z)$ and $\nu =-\frac{1}{2}+i\mu $
; $M^{2}=\frac{1}{4}+\mu ^{2}$%
\begin{equation}
G(n,n^{\prime })\sim Ae^{-i\mu L}+Be^{i\mu L}
\end{equation}%
So this Green function is unlikely to be the right starting point for the
quantum theory in the de Sitter space. The formula ( 5) is destroyed by the
analytic continuation because on a hyperboloid it becomes divergent.

A different Green function which does satisfy our requirement is given by%
\begin{equation}
G(n,n^{\prime})=Q_{\nu}(n\cdot n^{\prime})
\end{equation}
It contains only one exponential in the time-like asymptotics and has a
correct singularity at the coinciding points. But it also has an unexpected
singularity at the antipodal points defined by $(n\cdot n^{\prime})=-1.$%
Strange as it is, this singularity must appear if we define the propagator
as a sum over trajectories. Indeed, the antipodal points can be connected by
infinite number of geodesics. The functional integral includes the
integration over the zero mode describing the subspace of these geodesics.
If this subspace is non-compact, one expects and gets the divergence at the
antipodal points. We will discuss this singularity below.

Before discussing the physical consequences of this propagator, let us
analyze the AdS spaces. We can get them from a sphere by the continuation $%
n_{k}\Rightarrow in_{k}$ , so that in this case we have a two -sheeted
hyperboloid , $n_{0}^{2}-n_{k}^{2}=1.$The important difference from the
above is that we also have to change the sign of the metric on a sphere, $%
ds^{2}\Rightarrow-ds^{2}.$In this way we obtain the AdS space with the
euclidean signature (EAdS, or a real Lobachevsky space). Its metric is given
by%
\begin{equation}
ds^{2}=(dn_{k})^{2}-(dn_{0})^{2}=y^{-2}((dx)^{2}+(dy)^{2})=(d\rho)^{2}+%
\sinh^{2}\rho(d\omega)^{2}+\cosh^{2}\rho(dt)^{2}
\end{equation}
where we introduced for the future use two standard parametrizations of the
AdS. The scalar curvature $R=g^{\alpha\beta}R_{\alpha\beta}$ changes sign
under the above transformation $g_{\alpha\beta}\Rightarrow-g_{\alpha\beta}$
and hence the space has constant negative curvature. More interestingly,
under this transformation the covariant Laplacian also changes sign. Hence
we have the following relation between the Green functions%
\begin{equation}
G_{dS}(n,n^{\prime},M^{2})=G_{AdS}(n,n^{\prime},-M^{2})
\end{equation}
This formula implies that the positive mass in the de Sitter space
corresponds to a tachionic mass in AdS. That give us a first hint that the
dS space may be unstable. However these analytic relations don't tell us
which propagators are physical. This is determined by their spectral
properties.

\section{ Spectral features of various propagators}

We see that the propagators in the spaces of constant curvature are
analytically related. Let us begin to analyze their spectral properties with
the EAdS space. In the Poincare coordinates we have the following expression%
\begin{equation}
z=(n\cdot n^{\prime})=1+\frac{(x-x^{\prime})^{2}+(y-y^{\prime})^{2}}{%
2yy^{\prime}}=1+\frac{1}{\sin\sigma_{1}\sin\sigma_{2}}(\cosh(t_{1}-t_{2})-%
\cos(\sigma_{1}-\sigma_{2}))
\end{equation}
where we introduce variable $0$ $\leq\sigma\leq\pi$ $,$ $\cosh\rho=\frac {1}{%
\sin\sigma}.$ This variable appears when the Poincare upper half-plane is
conformally mapped onto an infinite strip. We see that since $z\geq1$ and we
must make a change $M^{2}\Rightarrow-M^{2}$ when passing from a sphere. The
only propagator which doesn't blow up in this space is 
\begin{equation}
G(n,n^{\prime})=Q_{-\frac{1}{2}+\mu}(z)
\end{equation}
where $M^{2}=-\frac{1}{4}+\mu^{2}.$ Its spectral decomposition is given by
the formula%
\begin{equation}
G(n,n^{\prime})=\int_{0}^{\infty}\lambda\tanh\pi\lambda\frac{P_{-\frac{1}{2}%
+i\lambda}(n\cdot n^{\prime})}{\lambda^{2}+\mu^{2}}
\end{equation}
This is a precise analogue of ( 4) in which the angular momentum $l$ is set
to be $-\frac{1}{2}+i\lambda.$It is also instructive to look at the massless
case, $\mu=\frac{1}{2}.$ Since the metric ( 10) is conformally flat, the
propagator in this case is just that of a free particle moving in the upper
half plane or , equivalently, in a strip with the Dirichlet boundary
conditions%
\begin{equation}
G(n,n^{\prime})=\log(\frac{z+1}{z-1})=\sum_{1}^{\infty}\frac{1}{k}e^{-k\mid
t\mid}\sin k\sigma_{1}\sin k\sigma_{2}
\end{equation}
If we take a Fourier transform with respect to $t,$the k-th term in this
expression will give $G_{k}\sim\frac{1}{\omega^{2}+k^{2}}.$The variable $t$
is conjugate to the dilatation operator. It is easy to see that in the case
of arbitrary mass, the answer will be 
\begin{equation}
G(\omega,\sigma_{1},\sigma_{2})=\sum_{k=0}^{\infty}\frac{\Phi_{k}(\sigma
_{1})\Phi_{k}(\sigma_{2})}{\omega^{2}+(k+\Delta)^{2}}
\end{equation}
Here $\Delta=\mu+\frac{1}{2}$ is the lowest dimension which is increased by
the raising operators ; the eigenfunctions are expressed through the
Legendre functions and form the space of the lowest weight unitary
representation of the symmetry group.

Our next task is to perform the Minkowskian continuation. It can be
accomplished by taking $t\Rightarrow it$ in the metric ( 10). It is also
possible to take one of the $x$ -s to be imaginary in Poincare's metric. The
fact that in the latter case the resulting coordinates don't cover the whole
space is of little consequence, since the propagators depend on the
invariant distances only. There are, nevertheless, a number of subtleties
which we have to discuss. The variable $z$ is given by the formula ( 12 ) in
which we have to replace $\cosh(t_{1}-t_{2})\Rightarrow\cos(t_{1}-t_{2}).$
As a result, now -$\infty\leq z\leq\infty.$Hence, we have to deal with
antipodal singularity at $z=-1$ in ( 13 ) and (15) . We might consider for a
moment to replace the $Q$ - function by the $P$ function which doesn't have
this singularity. However this choice is a disaster since the $P$ - function
blows up at infinity.

Another puzzle is related to the fact that any function of $n\cdot
n^{\prime} $ is periodic in $t$ variable. That means that they are defined
on a space with the closed time-like geodesics and not on the universal
covering space which we are primarily interested in. In other words these
propagators have singularities not only at $t=$ $t_{1}-t_{2}=0$ (if $%
\sigma_{1}=\sigma_{2} $) but also at $t_{1}-t_{2}=2\pi m.$

The resolution of these puzzles follow from the composition principle. It
dictates that while changing $\omega\Rightarrow i\omega$ in order to go the
Minkowskian AdS ( MAdS), we must take the following propagator%
\begin{equation}
G(\omega,\sigma_{1},\sigma_{2})=\sum_{k=0}^{\infty}\frac{\Phi_{k}(\sigma
_{1})\Phi_{k}(\sigma_{2})}{\omega^{2}-(k+\Delta)^{2}+i0}
\end{equation}
This formula gives the Feynman propagator for the universal covering space,
the $\widehat{MAdS}.$ If we are interested in the original space on which $t$
is a periodic variable, we must restrict $\omega$ to the integer values. If
we attempt to change the $i0$ prescription, say by taking a superposition of 
$+i0$ and $-i0$ terms for a propagator , we will destroy the composition,
since the convolution on the right hand side is divergent (the $\omega$
contour is pinched from the opposite sides).

In the coordinate space the function $G$ is not periodic in time, while
still having the form ( 13) away from the singularities. This happens
because the $i0$ prescriptions are different at $t=0$ and $t=2\pi m.$\qquad
This is clear already from the expression ( 15) since the factor $%
e^{ik\mid\tau\mid}$ is not periodic in $t.$ In the first case we have a
standard flat space singularity%
\begin{equation}
G\sim\log[(t^{2}-(\sigma_{1}-\sigma_{2})^{2}-i0]
\end{equation}
which, under the action of the Laplace operator, generates a delta function
at the coinciding points. At the same time near $t=2\pi m$ we have%
\begin{equation}
G\sim\log[(t-2\pi m+i0)^{2}-(\sigma_{1}-\sigma_{2})^{2}
\end{equation}
and there is no delta function at these points.

The singularities at $z=-1$ are physically necessary. They occur, according
to (12 ), when $\cos (t_{1}-t_{2})=\cos (\sigma _{1}+\sigma _{2}).$All
space-time is represented by a strip $0\leq \sigma \leq \pi .$ The above
relation means that the two points are connected by a light geodesic which
is reflected from $\infty $ odd number of times. Let us stress again that
our prescription requires to sum over all paths lying \textit{inside the
strip}. We do not allow them to go to outside . The boundaries of the strip
act as perfect mirrors. If the path is a null geodesics, we expect and get
singularities whenever $L(P_{xx^{\prime }})=0.$

Let us turn now to the case of the de Sitter space. All we have to do is to
interchange space and time in the AdS propagator and take $\mu\Rightarrow
i\mu$ (which inverts the sign of $M^{2}).$It is also important to remember
that in this case the space is periodic since we are dealing with the
one-sheeted hyperboloid. As a result, we get the following propagator%
\begin{equation}
G(n,n^{\prime})=Q_{-\frac{1}{2}+i\mu}(n\cdot n^{\prime}+i0)
\end{equation}
which satisfies the composition principle. At the same time the standard
propagator, proportional to the $P$ function blows up and the composition
integral diverges. It is instructive to analyze these propagators in the
Poincare coordinates , although they are geodesically incomplete. Their
advantage is simplicity. We have the following wave equation%
\begin{equation}
(\partial_{\tau}^{2}+p^{2}+\frac{M^{2}}{\tau^{2}})g_{p}(\tau,\tau^{\prime
})=\delta(\tau-\tau^{\prime})
\end{equation}
Here we made a Fourier transfom with respect to space variable $\sigma$ and
introduced the conjugate momentum $p$ . The resulting Green function is
given by $G_{p}(\tau,\tau^{\prime})=(\tau\tau^{\prime})^{\frac{1}{2}%
}g_{p}(\tau ,\tau^{\prime}).$The solution corresponding to the Bunch-Davies
vacuum is given by%
\begin{equation}
g_{p}(\tau,\tau^{\prime})\sim
H_{i\mu}^{(1)}(p\tau_{<})H_{i\mu}^{(2)}(p\tau_{>})
\end{equation}
which, in the case of massless particles is simply%
\begin{equation}
G_{p}(\tau,\tau^{\prime})\sim p^{-1}e^{-ip|\tau-\tau^{\prime}|}
\end{equation}
The prefactor $p^{-1}$would be $p^{-d}$ in d space dimension and is
responsible for the scale invariant spectrum of quantum fluctuations. It is
obviously the propagator obtained by the sum over paths lying in the \textit{%
whole plane.} At the same time, the Poincare coordinates are defined on the
upper half-plane. Therefore, the paths in the above propagator are allowed
to go to the \textit{\ \emph{wrong side of infinity}. }The propagator ( 20 )
has a different behavior. It includes paths which are reflected from
infinity ( $\tau=0),$but never go beyond it. This results in the Dirichlet
propagator in the massless case%
\begin{equation}
G_{p}(\tau,\tau^{\prime})\sim\frac{1}{p}[e^{-ip|\tau-\tau^{\prime}|}-e^{-ip(%
\tau+\tau^{\prime})}]
\end{equation}
while in the general case 
\begin{equation}
g_{p}(\tau,\tau^{\prime})\sim J_{i\mu}(p\tau_{<})H_{i\mu}(p\tau_{>})
\end{equation}
The relation of this propagator to the global one ( 20) follows from the
formula%
\begin{equation}
(\tau_{1}\tau_{2})^{\frac{1}{2}}\int dpJ_{i\mu}(p\tau_{<})H_{i\mu}(p\tau
_{>})e^{ipx}=Q_{-\frac{1}{2}+i\mu}(z)
\end{equation}
where $z$ is given by ( 12) with $y\Rightarrow i\tau.$

To understand it better, let us consider the spectral decomposition in the
global coordinates. In the de Sitter space these coordinates are obtained
from the usual polar coordinates, $ds^{2}=d\vartheta^{2}+\sin^{2}\vartheta
d\Omega^{2}$ on a sphere by the analytic continuation $\vartheta \Rightarrow%
\frac{\pi}{2}+it$ (where 0$\leq\vartheta\leq\pi$ is a polar angle, while the
global time -$\infty\leq t\leq\infty.$ On a sphere the eigenmodes are given
by $\phi_{m}\sim P_{l}^{m}(\cos\vartheta)$ (where $l$ is the angular
momentum and $m$ is a magnetic quantum number). As we make the above
substitution, we obtain the Bunch- Davies modes , $\phi_{m}\sim
P_{l}^{m}(i\sinh t),$where this time $l=-\frac{1}{2}+i\lambda.$These modes
are selected by the condition that they are\textit{\ non-singular at} $t=-%
\frac{i\pi}{2}$ (the south pole of the Euclidean sphere). The reason for
this requirement is based on the Hartle -Hawking geometry in which the
euclidean half-sphere ( presumably describing the tunneling creation of the
universe ) is joined with the Lorentzian half - hyperboloid. It is far from
obvious that this centaur is capable of living. We will see in the next
section that it is unstable in the sense discussed in the introduction..

Let us discuss other vacua, corresponding to the full hyperboloid.
Generally, the eigenmodes satisfy the Klein - Gordon equation%
\begin{equation}
\frac{\partial^{2}}{\partial t^{2}}\phi+(\lambda^{2}+\frac{(m-\frac{d-1}{2}%
)^{2}-\frac{1}{4}}{\cosh^{2}t})\phi=0
\end{equation}
(where we returned to $d+1$ dimensional case for the future references).
From this equation it is clear that the spectral decomposition of the Green
function must contain both discrete and continuous spectra. We can choose
the "in" state by $\phi\sim e^{-i\lambda t}$ as $t\rightarrow-\infty$ and ,
correspondingly the "out" state with the same asymptotic but at t$%
\rightarrow \infty.$These sets of modes ( we also denote them by ( $\pm)$ )
describe the vacua without particles at the past or future infinities. They
give rise to the " Dirichlet" propagator described above.

The "in" or "-" modes (here $d=1$ again, to simplify notations) are given by 
$\phi\sim Q_{l}^{m}(i\sinh t);l=-\frac{1}{2}+i\lambda``$ (these are the
standard Jost functions for this problem, which contain only one exponential
at $t\rightarrow-\infty$). This function must be understood as an analytic
continuation from $t>0.$The standard definition of the $Q$ function contains
a cut from -$\infty$ to 1. As $t$ becomes negative, we must go to the second
sheet through the cut. As a result we have $\phi\sim Q_{l}^{m}(i\sinh
t)+i\pi P_{l}^{m}(i\sinh t)$ for $t<0,$where the second term compensates the
jump across the cut, making $\phi$ an analytic function.These modes must be
complemented with the discrete states, which are given by the same formula
but with $0\leq$ $l\leq m-1$ (and being an integer). The "out" modes are
obtained by the $CPT$ reflection. The presence of the discrete states
reflects the fact that on a hyperboloid there are closed geodesics
(ellipses). The open geodesics (hyperbolae) correspond to the continuous
spectrum. In the case of the Hartle -Hawking geometry the regularity on the
south pole forces us to drop the discrete states, since the $Q$ function ,
unlike $P$ is singular there. To sum up, we have three types of modes
("vacua") , "in", "out" and E. The "E/out" propagator is given by the $Q$
function and satisfies the composition principle, the E/E propagator is
given by the $P$ function and does not ; all other propagators are
combinations of $P$ and $Q.$ We will also need the expression for the $Q$ -
propagator in $dS_{d+1}.$It is given by 
\begin{equation}
G(z)=const(1-z^{2})^{\frac{d-1}{4}}Q_{-\frac{1}{2}+i\mu}^{\frac{d-1}{2}}(z)
\end{equation}
where this time $M^{2}=\frac{d^{2}}{4}+\mu^{2}$ and as before $z=(n\cdot
n^{\prime}).$It is interesting to notice that for even $d$ this expression
reduces to the elementary functions. For example , for $d=2$ the answer is 
\begin{equation}
G=\frac{1}{4\pi i}\frac{e^{-i\mu l}}{\sinh l}
\end{equation}
where $l=\log(z+\sqrt{z^{2}-1})$ is the geodesic distance between the two
points. Our convention is that it is real for $z\geq1$ (time-like
separations ) , imaginary for $-1\leq z\leq1$ (moderate space-like
separations ) . For $z<-1$ the geodesic distance becomes complex. This is a
peculiar property of the $dS$ space - such points can't be connected by a
real geodesics.

\section{The composition principle vs. Hartle and Hawking}

In a certain sense our proposal for the propagators on eternal manifolds is
different from the Hartle-Hawking proposal for the wave functional. Indeed,
we could calculate the correlation function for the two points lying on the
equator of the hyperboloid in the following way. The d-dimensional geometry
of the equator is given (according to Hartle and Hawking) by the sum over
Euclidean geometries bounded by the equator. In the semi-classical
approximation this manifold is simply a hemisphere. So, for a free scalar
field we expect the wave functional of the form 
\begin{equation}
\Psi\lbrack\phi(\sigma)]\sim\exp\int d\sigma d\sigma^{\prime}D(\sigma
,\sigma^{\prime})\phi(\sigma)\phi(\sigma^{\prime})
\end{equation}
where the Poisson kernel $D(\sigma,\sigma^{\prime})=\partial_{\bot}%
\partial_{\bot}G(\sigma,\sigma^{\prime})$ is expressed through the normal
(with respect to the equator) derivatives of the Green function. In the case
of a sphere we can always decompose the functional integral in a following
way%
\begin{equation}
\int_{S}D\varphi=\int_{E}D\phi(\int_{S_{-}}D\varphi)(\int_{S_{+}}D\varphi)
\end{equation}
where $S_{\pm}$ are the hemispheres, $E$ is an equator, and the integrals in
the brackets are taken with the boundary condition that the bulk fields $%
\varphi$ approach a given value $\phi$ at the equator. We see that the
correlation function for the two points on the equator, calculated with
Hartle -Hawking wave functional is the same on a sphere and in the de Sitter
space. At the same time, the composition principle dictates that this
correlation is given by the $Q-$function in the de Sitter case and by the $P$%
- function on a sphere. The difference is that in one case the amplitudes
are given by the sum over paths lying in the full lorentzian manifold (a
hyperboloid in our case) , while in the other the one half of the
hyperboloid is replaced by an euclidean sphere.

The natural question to ask in this situation (which was already touched in
the introduction) is what is wrong with the standard approach in which we
postulate the Bunch- Davies (or the Hartle- Hawking) vacuum from the start
and to discard the composition principle .

To answer this question, let us assume for a moment that the Bunch -Davies
vacuum is the correct and stable one. That means that the $in$ and $out$
states coincide. Therefore, if we introduce some interaction, say $%
\lambda\varphi^{4}$, we can use the standard Feynman rules. This leads to
the incurable infrared divergence even for the massive particles . For
example, the first correction to the Green function has the form%
\begin{equation}
G^{(1)}(n,n^{\prime})\sim\lambda\int
dn_{1}G(n,n_{1})G(n_{1},n_{1})G(n_{1},n^{\prime})
\end{equation}
where we integrate over the hyperboloid. If we use the $P$ propagator, the
integrand contains the interference terms behaving as $z^{\nu}z^{-\nu-d}\sim
z^{-d}$(where $d$ is the number of space dimensions and $z=(n\cdot n_{1}$).
Since the measure $dn_{1}\sim z^{d-1}dz,$ we see that we have a logarithmic
divergence $\log z\sim L$ at large separations ( $L$ being a geodesic
distance) even for the massive particles. Clearly, in the higher orders we
will be getting stronger divergences.

To understand the meaning of this phenomenon, let us digress and consider as
an analogy the case of a free scalar field in the flat space-time but in a
state with arbitrary occupation numbers, $f(p),$ where $p$ is momentum. The
Green function in this state is given by%
\begin{equation}
G(p,t_{1},t_{2})=\frac{1}{2\omega (p)}[(1+f)e^{-i\omega
|t_{1}-t_{2}|}+fe^{i\omega |t_{1}-t_{2}|}]
\end{equation}%
where $\omega =\sqrt{p^{2}+m^{2}}$ . If we assume for a moment that this
state is stable in the above sense (equality of $in$ and $out)$ , the
Feynman perturbation theory leads us to a disaster, since in any loop
diagram the interference terms proportional to $f(1+f)$ will be \textit{%
linearly} infrared divergent due to the cancellation of the exponents. This
divergence has a simple meaning : the interaction is changing the occupation
numbers and creates a non-zero time derivative of $f(p).$In the case of weak
coupling this time derivative satisfies the Boltzmann equation (written here
for the $\varphi ^{3}$ case to simplify the notations)%
\begin{equation}
\frac{\partial f}{\partial t}\sim \int
[(1+f(p))f(p_{1})f(p_{2})-f(p)(1+f(p_{1})(1+f(p_{2})]\delta (\omega -\omega
_{1}-\omega _{2})\delta (p-p_{1}-p_{2})dp_{1}dp_{2}
\end{equation}%
The linear time divergence is absent only if the collision integral in (34 )
is zero. This is the steady state condition. To recapitulate, without
interaction we can have a stable state with arbitrary occupation numbers $%
f(p).$This stability is destroyed by any, however weak, interaction, unless
the occupation numbers are given by Bose/Fermi distributions.

Analogously, in our problem the linear divergence is a signal that stability
and the composition principle are violated. Indeed, if we use the doubled
space-time we can easily derive the formula%
\begin{equation}
\frac{\partial }{\partial m}G_{++}(n,n^{\prime })\sim \int
dn_{1}[G_{++}(n,n_{1})G_{++}(n_{1},n^{\prime
})-G_{+-}(n,n_{1})G_{-+}(n_{1},n^{\prime })]
\end{equation}%
In this integral the linear divergence cancels . The non-zero contribution
of the second term (violating the composition ) means that the vacuum we are
working with is unstable. The Bunch- Davies (E) vacuum \emph{fails the
eternity test }(we define this test as the vanishing of the second term in (
35) )\emph{. }

Let us investigate this condition in more details. The presence of the
spider diagrams imply that instead of the vacuum state we must look for an
excited states with the occupation numbers $\{f(p)\}$ over the Bunch -
Davies vacuum. In the small coupling limit these functions satisfy a
Boltzmann equation. The structure of this equation in this case is unusual,
since it includes particles creation from the vacuum. Leaving the detailed
analyzes for another work, we will only present these non-standard terms.
They have the form%
\begin{equation*}
\frac{\partial f}{\partial t}=\int dp_{1}dp_{2}\delta
(p+p_{1}+p_{2})|A|^{2}[(1+f(p))(1+f(p_{1}))(1+f(p_{2})]+O(f)
\end{equation*}%
where 
\begin{equation*}
A\sim \int_{0}^{\infty }\frac{d\tau }{\tau ^{d+1}}\tau ^{3d/2}H_{i\mu
}^{(1)}(p_{1}\tau )H_{i\mu }^{(1)}(p_{2}\tau )H_{i\mu }^{(1)}(p_{3}\tau )
\end{equation*}%
This term describes the vacuum decay since it is non-zero even when $f=0.$
The terms not explicitely displayed are the standard terms describing the
evolution of the one particle states. It seems plausible that the steady
state can be reached only when \emph{the gravity from the created excitation
screens the "anti-gravity" of the cosmological constant} leaving us with the
Friedman-like universe. I hope to discuss these fascinating questions
elsewhere.

Let us explore the non-interacing theory in more details. We expand the
partial wave of the field

\begin{equation}
\varphi =a\phi +a^{+}\phi ^{\ast }
\end{equation}%
where $\phi -$ s are some solutions of (27) and $a,a^{+}$ are the creation
and annihilation operators. The vacuum is defined by the $a|vac\rangle =0.$
Different choices of the solutions correspond to the different vacua. The
Green functions, $\langle vac|...|vac\rangle $ are given by%
\begin{equation}
G_{++}(t_{1},t_{2})=\phi (t_{>})\phi ^{\ast
}(t_{<});G_{+-}(t_{1},t_{2})=\phi (t_{2})\phi ^{\ast }(t_{1}),etc
\end{equation}%
Substituting these expressions ( 36 ) we find the eternity condition%
\begin{equation}
\int dt\phi ^{2}(t)=0
\end{equation}%
As we mentioned above, the modes corresponding to the E- vacuum are those
which remain finite at the south pole and have the form $\phi _{E}\sim
P_{\nu }^{m}(\cos \vartheta )$ , $\vartheta =\frac{\pi }{2}+it.$Let us
notice also that in terms of the conformal time $\tau $ ( $\tanh t=\cos \tau
)$ the euclidean south pole corresponds to $\tau \rightarrow i\infty .$That
explains why in the Poincare coordinates the E-vacuum requires the Hankel
function ( 22 ). The integral ( 38) can be written in the invariant form as $%
\int_{C}d\vartheta \sin \vartheta (P_{\nu }^{m}(\cos \vartheta ))^{2}$ (the
contour $C$ goes from zero to $\frac{\pi }{2}$ and then to $i\infty )$.
These integrals in can be expressed in terms of the scattering data. Namely,
let 
\begin{equation}
\phi (t)\rightarrow \alpha (\mu )e^{-i\mu t}+\beta (\mu )e^{i\mu
t},t\rightarrow \infty
\end{equation}%
Using the standard Wronskian identities, we find%
\begin{equation}
\int^{T}dt\phi ^{2}(t)=i\mu (\alpha \frac{\partial \beta }{\partial \mu }%
-\beta \frac{\partial \alpha }{\partial \mu })+2\alpha \beta T
\end{equation}%
We conclude that for a manifold to be eternal, we must have zero reflection
amplitude. Eternity is a rare thing.

We can also explore vacuum stability without doubling of the manifold. For
this purpose we introduce the $in/out$ and $in/E$ Green functions%
\begin{align}
G(n,n^{\prime}) & =\langle out|T\varphi(n)\varphi(n^{\prime})|in\rangle
\langle out|in\rangle^{-1} \\
G(n,n^{\prime}) & =\langle out|T\varphi(n)\varphi(n^{\prime})|E\rangle
\langle out|E\rangle^{-1}
\end{align}
where the time ordering is taken with the respect of the time component of $%
n.$In the first case we look at the development of the entire de Sitter
space - the full hyperboloid. In the second case we are dealing with the
centaur - Euclidean hemisphere joined with the Lorentzian hyperboloid along
the equator. As we noticed before, they are given by the $Q$ propagator, the
only one satisfying the composition principle.This is all we need, we don't
have to know what are the $in$ and $out$ states. Using the formula for the
effective action, $e^{iW}=\langle out|in\rangle$ we find the standard
formula 
\begin{equation}
\frac{\partial W}{\partial\mu}\sim\int dnG(n,n)
\end{equation}
Of course the Green function at the coinciding points is infinite, but its
imaginary part is finite and determines the imaginary part of the effective
action (which will have also a trivial infinity - the total volume of the dS
space, since the integrand is independent of $n$). The result is 
\begin{equation}
\frac{\partial}{\partial\mu}\func{Im}W\sim(Vol)\func{Im}Q_{\nu}(1)
\end{equation}
(where $(Vol)$ is the infinite volume of the whole space ). Using the formula%
\begin{equation}
Q_{-\frac{1}{2}+i\mu}(z+i0)-Q_{-\frac{1}{2}-i\mu}(z-i0)=i\pi(\tanh\pi
\mu-1)P_{-\frac{1}{2}+i\mu}(z)
\end{equation}
we find 
\begin{equation}
\func{Im}W\sim(Vol)\log(1+e^{-2\pi\mu})
\end{equation}

This method of finding the vacuum decay mimics the derivation of Schwinger
of pair production in the strong electric fields. It is interesting that the
effective action can also be expressed in terms of the scattering data. Let
us consider the effective action $W_{m}$ for the given angular momentum $m.$%
The partial $in/out$ Green function is given by 
\begin{equation}
G_{m}(t,t^{\prime})=\frac{1}{2\mu\alpha_{m}(\mu)}\phi_{m}(t_{>})\chi_{m}^{%
\ast}(t_{<})
\end{equation}
where $\varphi$ and $\chi$ are the Jost functions for the eq. ( 27) ( having
the asymptotics $e^{-i\mu t}$ at $t\rightarrow\pm\infty$ respectively ,
while $\frac{1}{\alpha}$ is the transmission coefficient). The partial
effective action satisfies the identity%
\begin{equation}
i\frac{\partial W_{m}}{\partial\mu^{2}}=\frac{1}{2}\int dtG_{m}(t,t)\sim 
\frac{1}{\alpha_{m}}\int dt\phi_{m}(t)\chi_{m}^{\ast}(t)\sim\frac{\partial
\log\alpha_{m}(\mu)}{\partial\mu}
\end{equation}
The last integral was again evaluated using the Wronskians. Collecting all
factors , we find a curious formula 
\begin{equation}
e^{iW_{m}}=\frac{1}{\sqrt{\alpha_{m}}}=\sqrt{T_{m}}
\end{equation}
where $T_{m}$ is the transmission amplitude. We see once again that the
non-vanishing imaginary part of the effective action is related to the
presence of reflection. The reflection amplitude for the equation (27) can
be extracted from the reader's favorite text book on quantum mechanics ( the
Landau/ Lifshits in my case). But one must be careful in comparing the
partial wave actions with the complete one ( 44), since the summation on $m$
diverges and can easily give a wrong result ( instead of the infinite
invariant volume)

We come to the conclusion that the de Sitter space of even dimensionality is 
\emph{intrinsically unstable}. Whatever we do, we can't build the eternal
space. In the case of the odd dimensions we don't get the imaginary part of
the action. The reason for this can be traced to the equation (27). If $d$
is odd , the potential in this Shrodinger equation can be written as $\frac{%
n(n+1)}{\cosh^{2}t}$ with integer $n$ . This is a soliton of the KdV
equation which gives \textit{reflectionless potential }( the vanishing of
the reflection coefficient was noticed by the explicit calculations in \cite%
{stro} ).

As a side remark let us notice that in the case of general FRW spaces with
the scale factor $a(t)$ one may try to look at the more general
reflectionless potentials, corresponding to the separated solitons, in an
attempt to find eternal backgrounds. This relation between eternal manifolds
and completely integrable systems seems quite intriguing.

Returning back to the odd dimensional spaces we come to the following
conclusion. If we consider a full de Sitter space, with the past and future
infinities it seems that at least in one loop approximation we have a stable
manifold. However, if we consider the Hartle - Hawking geometry, the
amplitude $\beta\neq0$ and the stability is lost.

As another side remark, let us notice that in the case of AdS spaces the
above analyzes indicates that the simply connected global AdS is stable, but
the one with the closed time-like geodesics is not (this is easily seen from
the propagators (17 ) in which $\omega$ becomes discrete in the multiply
connected case ). Such geodesics seem to destabilize a manifold in general
by giving an imaginary part to the $in/out$ propagators in the formula ( 1).

\section{The infrared effects in the de Sitter space}

In this section we will discuss infrared interactions. We already saw how
significant they are in a number of examples. It is desirable to have a
general estimate of their strength. In the flat space such estimates are
well known and very important - they led to the notion of relevant and
irrelevant operators. In \cite{p81} I tried to evaluate the infrared
corrections to the cosmological constant in the Euclidean signature, while
in the later work \cite{wood} and \cite{weinb} a number of results were
obtained in the in-in formalism.

Let us remember the infrared situation in the flat space, by considering the
amplitude for two interacting paths ( with $\varphi^{4}$ interaction). The
relevance of this interaction is determined by the probability for these two
Brownian paths to intersect. The standard estimate is to compare the
propagation without interaction which contributes $G^{2}(R)$ (where $G\sim%
\frac{1}{R^{D-2}}$ is the Green function and $R$ is a typical distance) with
the interactiion term which gives $G^{4}(R)R^{D}$ (where the second factor
is the volume in which the interaction takes place). From here one finds the
critical dimension $D_{cr}=4.$

Let us analyze the de Sitter case in a similar manner. The Bunch - Davies
propagator has the following asymptotic behavior 
\begin{equation}
G(z-i0)\rightarrow z^{-\frac{d}{2}}(A(\mu)z^{\mu}+A(-\mu)z^{-\mu})
\end{equation}
where $z=(n_{1}n_{2})$ and $\mu=\sqrt{\frac{d^{2}}{4}-M^{2}}.$ We have to
distinguish the cases of the \textit{light particles} with $M^{2}\leq \frac{%
d^{2}}{4}$and the heavy ones with the opposite inequality. In the previous
sections we dealt mostly with the heavy particles for which we replaced $%
\mu\Rightarrow i\mu.$ Here we will be interested in both cases. Let us
consider the $\varphi^{N}$ interaction and begin with the Feynman
perturbation theory. Its significance is measured by the amplitude%
\begin{equation}
F(n_{1}...n_{N})\sim\int(dn)G(nn_{1})...G(nn_{N})
\end{equation}
To evaluate convergence of this integral we use the first term in (50 ) and
the fact that at large $z$ (the infrared limit) the measure $(dn)\sim
z^{d-1}dz.$

We see that for the sufficiently light particles the integral has a power
like divergence. The condition for it is given by%
\begin{equation}
N\mu\geq\frac{N-2}{2}d
\end{equation}
For the $\varphi^{4}$ interaction this gives the condition $M^{2}\leq \frac{%
3d^{2}}{16}.$ The massless case, $\mu=\frac{d}{2}$ , is always IR divergent.
Because of the $-i0$ prescription, in general these divergent terms have
imaginary parts (which may cancel in some special cases).

If the mass is large enough for the above integral to converge, it is easy
to see that it is dominated by the region $(n_{i}n_{j})\sim (n_{k}n)^{2}\gg
1.$In this case an easy estimate shows that the interaction is \textit{%
always marginal }. This can be seen in the Poincare coordinates in which the
integral (51) takes the form%
\begin{equation}
F\sim (\tau _{1}...\tau _{N})^{\frac{d}{2}-\mu }\int \frac{d^{d}xd\tau }{%
\tau ^{d+1}}\frac{\tau _{{}}^{N(\frac{d}{2}-\mu )}}{\dprod
((x-x_{k})^{2}-(\tau -\tau _{k})^{2})^{\frac{d}{2}-\mu }}
\end{equation}%
In this form ( 52) is just the divergence (convergence) condition at $\tau
\rightarrow 0.$As we compare the value of $F$ at small $\tau _{k}$ with the
non-interacting $N$ - point function, we obtain the above result for the
convergent case.

In order to study back reaction we need to know various polarization
operators. The simplest one appears if we add to the Lagrangian the a term $%
\int(dn)A(n)J(n)$ where $J(n)$=$\varphi^{2}(n).$As usual we define the back
reaction (for the massless case )from the relation%
\begin{equation}
\delta J(n)=\int(dn^{\prime})\Pi(n,n^{\prime})A(n^{\prime})=\int(dn^{\prime
})G^{2}(nn^{\prime})A(n^{\prime})\sim\int(dn^{\prime})\log^{2}(nn^{\prime
}-i0)A(n^{\prime})
\end{equation}

In the case of $in/in$ formalism the structure of the infrared divergences
is slightly different because of the partial cancellations between the $%
(\pm) $ Green functions (see also \cite{weinb} ). To evaluate them we need
the relations 
\begin{align}
G_{++}(n,n^{\prime}) & =g(nn^{\prime}-i0);G_{+-}(n,n^{\prime})=g(nn^{\prime
}+i\epsilon sgn(n_{0}-n_{0}^{\prime})) \\
G_{--}(n,n^{\prime}) & =g(nn^{\prime}+i0);G_{-+}(n,n^{\prime})=g(nn^{\prime
}-i\epsilon sgn(n_{0}-n_{0}^{\prime}))
\end{align}
The back reaction formula (54) is replaced by%
\begin{equation}
\delta
J(n)=\int(dn^{\prime})[G_{++}^{2}(n,n^{\prime})-G_{+-}^{2}(n,n^{\prime
})]A(n^{\prime})
\end{equation}
This time the integration goes over the region inside the past light cone (
outside the cone $G_{++}=G_{+-}$ and the integrand is zero). Nevertheless we
still have the long range effect since $\log^{2}(z-i0)-\log^{2}(z+i0)=2\pi
i\log|z|.$ Thus%
\begin{equation}
\delta J(n)\propto\int(dn^{\prime})\theta(n_{0}-n_{0}^{\prime})\theta
(nn^{\prime}-1)\log(nn^{\prime})A(n^{\prime})
\end{equation}

The formula ( 58) must be substituted into the equations of motion for the
field $A.$The resulting effective equations can't be derived from any action
principle, since the kernel in ( 58) is not symmetric with respect to $n$
and $n^{\prime}$ . The long range correlations should generate the
Jeans-like instabilities. However, to make the discussion realistic, we must
replace the scalar field with the gravitational fluctuations, fix the gauge
and account for the tensor structure. This may change the answer. We leave
this for the future work.

The contribution of the heavy particles ( for which we have to change $%
\mu\Rightarrow i\mu$ ) to the polarization operator is dominated by the
interference term. As a result , in the Feynman case , we get the kernel $%
\sim(nn^{\prime})^{-d}.$ This creates a logarithmic divergence, which we
already discussed in connection with the composition principle

\section{The gauge/strings duality}

Let us discuss now the gauge theory dual of the de Sitter space. It was
pointed out in \cite{p06} that the natural candidate for it is a gauge
theory with the complex coupling constant. We can now present more arguments
in favor of this assertion.

Gauge/ string duality is usually understood as a relation between the gauge
theory correlators and the string theory correlators in D+1 dimension placed
at infinity (which is D-dimensional as it should). While it is often
convenient, this point of view may cause trouble in the cases when the
manifold has a complicated infinity or no infinity at all. Perhaps a more
general construction is to consider the isomorphism between the vertex
operators of string theory ( a set of primary operators with the world sheet
dimensions ( $1,1)$ ) and a set of the field theory operators. In other
words we consider a 2d CFT on the world sheet with a critical central charge
( the Liouville field is assumed to be included in the menu). We couple it
to the world sheet gravity , which ( in this case ) amounts to selecting the
above primaries and integrating them over the world sheet, forming the
vertex operators. The resulting objects are identified with the colorless
field theory operators and the world sheet OPE are transplanted to the
space-time.

The derivation of the gauge/ strings duality from the first principles is a
fundamental and unsolved problem. Still, there are a number of indirect
qualitative arguments helping to find the duality in each particular case.
The simplest one runs as following. Suppose that we have a string theory in
the background $ds^{2}=d\varphi^{2}+a^{2}(\varphi)dx^{2}$ , where $x$ are
the coordinates of the space-time in which a gauge theory is located, while
the $\varphi$ is the Liouville direction. Let us place the Wilson loop at
the value $\varphi=\varphi_{\ast}$ such that $a(\varphi_{\ast})=\infty.$This
Wilson loop defines the open string amplitudes. The slopes of closed and
open strings are different. In this case we have $\alpha_{open}^{\prime}\sim
a^{-2}(\varphi_{\ast})\alpha_{closed}^{\prime}\rightarrow0.$ But in the zero
slope limit only the massless states survive, giving the gauge theory. The
closed string still has infinite number of states corresponding to the gauge
invariant operators, while the open string has finite number of massless
states (gluons etc.) The key point of the argument is that the infinite blue
shift sends all massive modes of the open string to infinity.

We can now apply this argument to two possible cases. The first case is the
centaur geometry (half -sphere, half- hyperboloid). The topology in this
case is the same as in the $AdS,$ which has the one component infinity ,
conformally equivalent to the flat space. Once again, the $a$ factor
provides an infinite blue shift and we can expect that there is a gauge
theory in d- dimensions, describing the d+1 dimensional strings in the
centaur background. The single trace operator of the gauge theory must be
equal to the vertex operators, giving the usual relation for the generating
functions%
\begin{equation}
\log\langle\exp i\int dx\sum_{n}h_{n}(x)TrO_{n}\rangle=iW[h_{n}(x,t)]
\end{equation}
where the $TrO_{n}$ are the gauge-invariant operators, while $W$ is the
effective action in the $h-$background for the strings on the centaur. The
right hand side is the generating function for the vertex operators. This
relation means that we can identify gauge and vertex operators in a
following way (in the Poincare coordinates)%
\begin{equation}
TrO_{n}(x)=\int d^{2}\xi V_{n}(x+x(\xi),y(\xi))
\end{equation}
where the integration goes along the world sheet and the $V_{n}(\xi)-$ s are
the primary operators of the string sigma model which satisfy the on-shell
condition $(L_{0}-1)V_{n}=0.$In the low curvature limit this condition
reduces to the Klein- Gordon equation. The expectation values $\langle
V_{n_{1}}...V_{n_{k}}\rangle$ are given by the sum of tree diagrams in
space-time with the set of massive and massless string states and the legs
at infinity. In the $AdS$ case the propagators in this diagrams are given by
the $Q_{-\frac{1}{2}+\mu}(z)$ function . To pass to the case of the positive
curvature , all we have to do is to change $\mu\Rightarrow i\mu$ . As a
result we obtain the same tree diagrams but with the propagators $Q_{-\frac{1%
}{2}+i\mu}.$As we have learned, these propagators correspond to taking
matrix elements $\langle out|...|E\rangle$ in the centaur geometry. The new
piece of information in these formulae is the statement that the analytic
continuation from $AdS$ to $dS$ (or , more precisely, to the centaur) leads
to the $out/E$ matrix elements, see also \cite{malda} ).

Let us discuss first the $\mathcal{N=}4$ gauge theory. In the strong
coupling limit the change $\mu\Rightarrow i\mu$ corresponds to the change of
the gauge coupling $\sqrt{\lambda}=\sqrt{g_{YM}^{2}N}\Rightarrow-\sqrt{%
\lambda}.$This is obviously a non-unitary conformal theory with some of the
anomalous dimensions being negative or complex, giving another manifestation
of the intrinsic instability of the de Sitter space. Strictly speaking, in
this example we have not only the de Sitter background but also the RR
fields which become imaginary under the analytic continuation. Gauge
theories with the complex coupling are expected to be unstable and so are
the corresponding de Sitter spaces. Apart from other things, the change of
the sign of $\sqrt{\lambda}$ at large $\lambda$ changes the sign of the
Coulomb interaction of two charges. In this case one expects Dyson's
instability, which potentially can terminate de Sitter's inflation.

So far we discussed the continuation of the maximally supersymmetric case,
which is the easiest but not terribly realistic. In the next section we look
at the more general situation.

\section{De Sitter sigma model}

To discuss string theory we need a conformal sigma model on the world sheet
with the de Sitter target space. Let us begin with a well studied case of a
sphere. It will be convenient for our discussion to discretize one of the
world-sheet directions, keeping the other continuous. In this case the
hamiltonian of the model can be written as 
\begin{equation}
H=\alpha_{0}\sum_{x}l_{x}^{2}+\frac{1}{\alpha_{0}}\sum_{x}(n_{x}-n_{x+1})^{2}
\end{equation}
where $\alpha_{0}$ is the bare coupling constant , $n$ is a unit vector, and 
$l$ is the corresponding angular momentum operator. This model is not
conformal and developes a mass gap. Let us remember how to see this (
without using the available exact solution of this model). First one
calculates the one loop beta function and find that this model is
asymptotically free. That means that the coupling grows as we go to the
infrared and if the higher orders don't stop it, leads to the mass gap $%
m^{2}\sim\exp(-\frac {const}{\alpha_{0}})$ at small coupling. If, on the
other hand there is a zero of the beta function at some finite value of $%
\alpha$ , the theory will have no mass gap and will be conformal. To choose
between the possibilities we look at the strong coupling limit of ( 61) .
The second term can be neglected and the model reduces to a collection of
uncoupled rotators. The lowest excitation is obtained by taking $l=1$ at
some site and has a mass gap $\sim\alpha_{0}$ at large $\alpha.$ We conclude
that the beta function has no zeroes and the theory is massive for all
couplings. Of course it is possible that the beta function would have two
zeroes, but this seems unlikely and indeed, according to the exact solution,
doesn't happen.

We see that the outcome depends on two factors - first, the sign of the
one-loop beta function which is determined by the sign of the curvature of
the target space and the second - compactness/ non-compactness of this
space. The compactness in the above example was responsible for the
discreteness of the spectrum and the possibility to drop the last term in (
61).

Let us now consider generalizations of this method for dS and AdS spaces and
their cosets. Perhaps the simplest case is the one with constant negative
curvature but compact target space. This can be realized in two ways. Either
one considers the so called $O(n)$ models with $n<2$ (the curvature of an $%
n- $sphere $\sim n-2)$, or one factors AdS space by a discrete subgroup with
the compact fundamental domain. In both cases we have positive beta function
at the week coupling and the discrete spectrum in the strong coupling limit.
Therefore we expect the beta function\textit{\ to have a zero at some
intermediate coupling}. In the case of the $O(n)$ models these conformal
fixed points are well known - they are simply and explicitly described by
the minimal models. In the case of the compact cosets the conformal theory
is still unknown.

As we return to the de Sitter space, we notice first of all that its beta
function coincides with that of a sphere in all orders of perturbation
theory. This follows from the fact that the transformation $n_{0}\Rightarrow
in_{0}$ taking us from the sphere to the dS space doesn't change the any
Feynman diagram. Nevertheless these two theories are very different. We have
already mentioned that the angular momentum in the dS space (or in centaur
geometry takes the values $l=-\frac{1}{2}+i\lambda$ where $\lambda$ is real.
That means that the spectrum of kinetic energy in the strong coupling limit
is continuous and we expect no mass gap. This is a conjecture. What we said
so far doesn't really prove it because in the non-compact case we can't
simply drop the potential energy, since $n$ can become large. Some
variational estimates show that by taking the slow varying $n$ we can
suppress the potential energy and end up with the gapless spectrum.
Intuitively the continuous spectrum appears because the de Sitter space is
non-compact, unlike a sphere. It is highly desirable to have a proof of this
statement.

If we believe this conjecture, we must conclude that the dS sigma model has
an infrared fixed point which resolves the conflict between the negative
beta function at the weak coupling and the gapless phase at the strong.
Generally speaking the central charge at this fixed point is smaller the the
critical one. However, adding real RR fields to the background allows to
adjust it to the critical value.

So we expect that the de Sitter space without any supersymmetry can be
described by a string sigma model. This model has a fixed point determined
by the zero of the beta function. That means that the curvature of the AdS
is fixed. The dual gauge theory must also be at the fixed point. At the
moment we do not have explicit description of these fixed point. We will
make the following conjecture about its origin.

In '74 t'Hooft conjectured that planar diagrams become dense and can be
described by a string's world sheet. In many recent papers it is stated that
AdS/ CFT is a realization of this idea. This statement is wrong because in
the usual gauge/ string duality Feynman's diagrams don't become dense at
all. The origin of strings in this case are electric flux lines , something
quite different from the propagators.

The planar diagrams have a finite radius of convergence and at some complex
coupling constant really become dense. It is tempting to relate that to the
fixed point of the de Sitter sigma model described above. If this is
correct, the " t'Hooft string" describes gauge theory at the critical
complex coupling and lives in the de Sitter space. However, at present we
don't have the necessary tools to check this conjecture.

So far we discussed the Hartle - Hawking geometry. Next, it is natural to
ask what is the gauge theory dual in the complete de Sitter space. The main
feature in this case is the existence of two infinities, past and future. It
is natural to conjecture that in this case we are dealing with two
interacting gauge theories living at these far away locations. The nature of
this duality and the possible forms of the trace-trace interactions remains
to be clarified.

I would like to thank T. Damour for our conversations over many years if not
decades about infrared effects in cosmology. I am very grateful to J.
Maldacena for sharing his (different ) views on dS space. Discussions with
I. Klebanov and N. Nekrasov were , as always, very useful.

Finally, my deep gratitude goes to D. Makogonenko for constant
encouragement, advice and moral support.

This work was partially supported by the NSF grant 0243680. Any opinions,
findings and conclusions or recommendations expressed in this material are
those of the authors and do not necessarly reflect the views of the National
Science Foundation.

\end{document}